\documentclass[12pt,a4paper,dvips]{article}
\usepackage{a4p}
\usepackage{cite,mcite}
\usepackage{graphicx}
\usepackage{physics }
\usepackage{l3_title,ifthen}
\usepackage{rotating}
\usepackage{epsfig}

\journalname{Phys. Lett. B}
\date{May 15, 2002}

\preprint{2002-034}

\usepackage{Lep}
\Lep{2}

\newcommand{\eeto}{\ensuremath{\rm e^+e^- \rightarrow}~}

\newcommand{\qqgg}{\ensuremath{\rm q\bar{q}\gamma\gamma}}
\newcommand{\Zgg}{\ensuremath{\rm Z\gamma\gamma }~}

\newlength{\capindent}
\newlength{\capwidth}
\newlength{\figwidth}
\newcommand{\icaption}[2][!*!,!]{\hspace*{\capindent}%
  \begin{minipage}{\capwidth}
    \ifthenelse{\equal{#1}{!*!,!}}%
      {\caption{#2}}%
      {\caption[#1]{#2}}
  \end{minipage}}
\begin{document}
\setlength{\unitlength}{1mm}
\begin{titlepage}
\title{The \boldmath{${\epem\ra\Zo\gamma\gamma}\rightarrow \qqgg $}
Reaction at LEP\\ and Constraints on\\ Anomalous Quartic Gauge Boson Couplings}
\author{The L3 Collaboration}
\begin{abstract}

 The cross section of the process
$\epem\ra\Zo\gamma\gamma\rightarrow\qqgg$ is measured with
215\,pb$^{-1}$ of data collected with the L3 detector during the final
LEP run at centre-of-mass energies around $205\GeV$ and $207\GeV$. No
deviation from the Standard Model expectation is observed.  The full
data sample of 713\,pb$^{-1}$, collected above the Z resonance, is
used to constrain the coefficients of anomalous quartic gauge boson
couplings to:
\begin{center}
  \begin{tabular}{ccl}
    $-0.02\GeV^{-2} <$&$ a_0/\Lambda^2 $&$< 0.03\GeV^{-2}$ and\\
    $-0.07\GeV^{-2} <$&$ a_c/\Lambda^2 $&$< 0.05\GeV^{-2}$,\\
  \end{tabular}
\end{center}
 at 95\% confidence level.

\end{abstract}

\submitted

\end{titlepage}

\section*{Introduction}

High energy $\epem$ collisions offer a unique environment to unveil
the structure of the couplings between gauge bosons. Extensive studies
of boson pair-production are performed to probe triple vertices of
neutral and charged bosons.  Results were recently reported on the
investigation of triple boson production through the reactions
$\epem\ra\Wp\Wm\gamma$~\cite{opalwwg,l3wwg} and
$\epem\ra\Zo\gamma\gamma$ \cite{l3pap,l3pap2}. These processes give
access to possible anomalous Quartic Gauge boson Couplings (QGCs).
 
\begin{figure}[p]
  \begin{center}
      \mbox{\includegraphics[width=\figwidth]{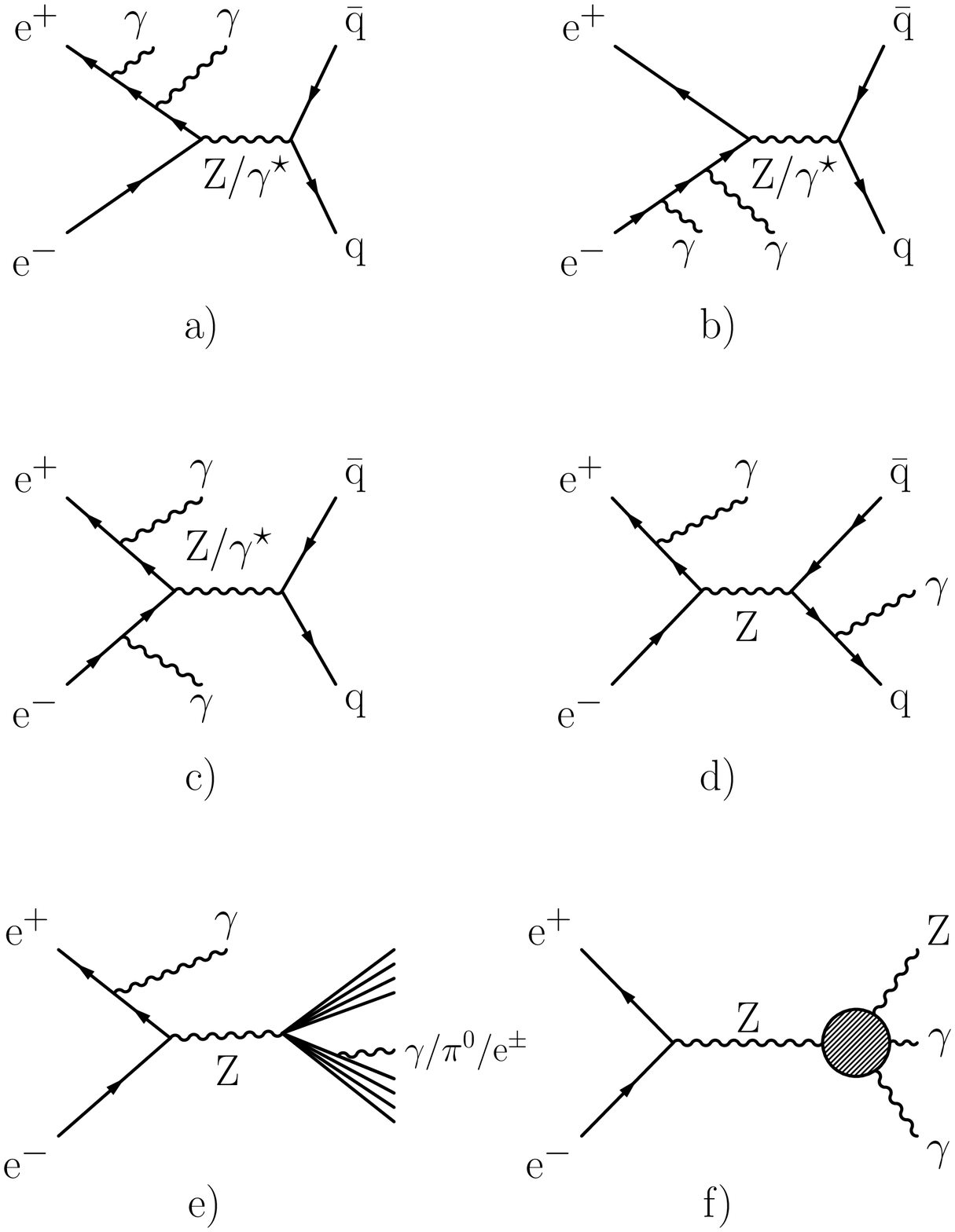}}\vspace{4cm}
      \icaption{Representative diagrams of, a)--c), the Standard Model
        contribution to the \eeto \Zgg signal and the ``non-resonant''
        background, d), the background from direct radiation of a
        photon from the quarks, e), the background from photons,
        misidentified electrons or unresolved $\pi^0$'s originating
        from hadrons and, f), the anomalous QGC diagram.
     \label{fig:0}}
  \end{center}
\end{figure}

Figures~\ref{fig:0}a$-$c display three of the six Standard Model diagrams
that describe the $\epem\ra\Zo\gamma\gamma$ process with the radiation
of photons from the incoming electrons. This
process is studied exploiting the high branching fraction of the Z
boson decay into hadrons. The $\epem\ra\Zo\gamma\gamma\ra\qqgg$ signal
is defined~\cite{l3pap2} by phase-space requirements on the energies
$E_{\gamma}$ and angles $\theta_\gamma$ of the two photons, on the
propagator mass $\sqrt{s'}$ and on the angle $\theta_{\rm \gamma q}$
between each photon and the nearest quark:
\begin{equation}
E_{\gamma}> 5 \GeV,\,\,\,\,\,\,
|\cos{\theta_{\gamma}}|< 0.97,\,\,\,\,\,\,
|\sqrt{s'}-m_{\rm Z}|< 2 \Gamma_{\rm Z},\,\,\,{\rm and}\,\,\,
\cos{\theta_{\rm \gamma q}}< 0.98,
\label{eq:sign}
\end{equation}
where $m_{\rm Z}$ and $\Gamma_{\rm Z}$ are the Z boson mass and
width.
Events with hadrons and initial state photons falling outside the
signal definition cuts are referred to as ``non-resonant'' background.

A single initial state radiation photon can also lower the effective
centre-of-mass energy of the $\epem$ collision to around $m_{\rm Z}$. This
photon can be mistaken for the most energetic photon of the signal and
two sources can then mimic the least energetic photon: the direct
radiation of photons from the quarks, or photons originating from
hadronic decays, misidentified electrons or unresolved $\pi^0$'s.
These background processes are depicted in Figures~\ref{fig:0}d
and~\ref{fig:0}e, respectively. 

In the Standard Model, the \Zgg production  via QGCs is forbidden at tree level.   Possible contributions of anomalous QGCs, through
the diagram sketched in Figure~\ref{fig:0}f, are described by two
terms of dimension-six in an effective Lagrangian~\cite{bb1,sw}:
\begin{eqnarray*}
{\cal L}^0_6 & = &  -{\pi\alpha \over 4\Lambda^2} a_0 F_{\mu\nu}F^{\mu\nu}
\vec{W}_\rho\cdot\vec{W}^\rho\\
{\cal L}^c_6 & = &  -{\pi\alpha \over 4\Lambda^2} a_c F_{\mu\rho}F^{\mu\sigma}
\vec{W}^\rho\cdot\vec{W}_\sigma,
\end{eqnarray*}
where $\alpha$ is the fine structure constant, $F_{\mu\nu}$ is the
photon field and $\vec{W}_\sigma$ is the weak
boson field.  The parameters $a_0$ and $a_c$ describe the strength of
the QGCs and $\Lambda$ represents the scale of the New
Physics responsible for these anomalous contributions. In the Standard
Model, $a_0 = a_c = 0$.  Experimental limits on QGCs were derived from
studies of the $\epem\ra\Wp\Wm\gamma$
process~\cite{opalwwg,l3wwg}. However, the $a_0$ and $a_c$
couplings might be different in the $\epem\ra\Zo\gamma\gamma$
case. Alternative parametrisations can be found in
References~\citen{bb2} and~\citen{racoon}.  Indirect bounds on QGCs
were extracted in Reference~\citen{eboli} using Z pole data.

\section*{Data Analysis}

Reference~\citen{l3pap2} describes the analysis of the
$\epem\ra\Zo\gamma\gamma\ra\qqgg$ process with $497.6\,{\rm pb}^{-1}$ of
data collected by the L3
detector~\cite{l3_00} at
LEP at centre-of-mass energies, $\sqrt{s}$, between 130 and 202
\GeV. This Letter details the equivalent findings from the final LEP
run, when the machine was operated at $\sqrt{s} = 200-209 \GeV$. These
data are grouped in two energy bins around average
$\sqrt{s}$ values of $204.8\GeV$ and $206.6\GeV$, respectively
corresponding to  integrated luminosities of $77.4\,{\rm pb}^{-1}$ and
$137.9\,{\rm pb}^{-1}$.

The signal and the ``non-resonant'' background are
described with the KK2f Monte Carlo program\cite{KK2f}, which takes
into account the interference of diagrams with initial and final state
photons. It is interfaced with the JETSET~\cite{pythia} program for
the simulation of hadronisation.

Other backgrounds are generated with the Monte Carlo
programs PYTHIA \cite{pythia} ($\rm e^+ e^- \rightarrow Z \epem$ and
$\rm e^+ e^- \rightarrow ZZ$), KORALZ\,~\cite{koralz} ($\rm e^+ e^-
\rightarrow \tau^+ \tau^- (\gamma)$), PHOJET\,~\cite{phojet} ($\rm e^+
e^- \rightarrow e^+ e^-$ hadrons) and KORALW\,~\cite{koralw} for
$\Wp\Wm$ production except for the $\rm e\nu_\e q\bar q'$ final
states, generated with EXCALIBUR~\cite{exca}.  The L3 detector
response is simulated using the GEANT~\cite{geant} and
GHEISHA~\cite{gheisha} programs, which model the effects of energy
loss, multiple scattering and showering in the detector. Time
dependent detector inefficiencies, as monitored during data taking
periods, are also simulated

Candidates for the $\epem\ra\Zo\gamma\gamma\ra\qqbar\gamma\gamma$
process are longitudinally and transversely balanced hadronic events
with two isolated photons with reconstructed energy above $5\GeV$,
detected in a polar angle range $|\cos{\theta}|<0.97$.  The invariant
mass of the reconstructed hadronic system, $M_{\rm q\bar{q}}$, is
required to be consistent with $m_{\rm Z}$: $74\GeV < M_{\rm
q\bar{q}}< 111\GeV$.

The main background after these requirements is due to the
``non-resonant'' production of two photons and a hadronic system. The
relativistic velocity $\beta_{\rm Z} = p_{\rm Z}/E_{\rm Z}$ of the Z
candidate is calculated from the kinematics of the observed photons,
assuming its mass to be $m_{\rm Z}$. As shown in Figure~2a,
$\beta_{\rm Z}$ is larger for part of these background events than for
the signal. Requiring $\beta_{\rm Z}<0.73$ rejects half of this background.

\begin{figure}[p]
  \begin{center}
    \begin{tabular}{cc}
      \mbox{\includegraphics[width=.5\figwidth]{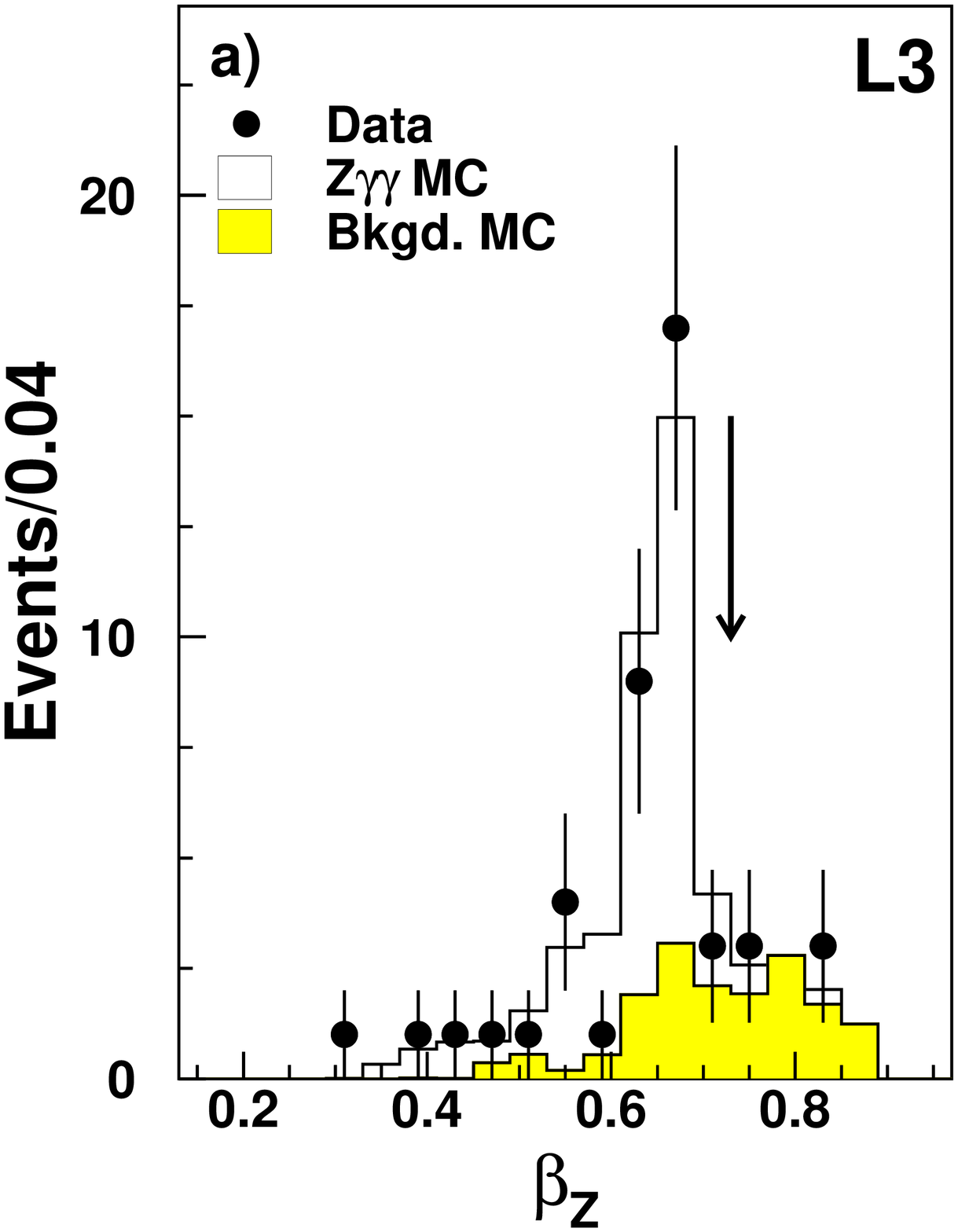}} &
      \mbox{\includegraphics[width=.5\figwidth]{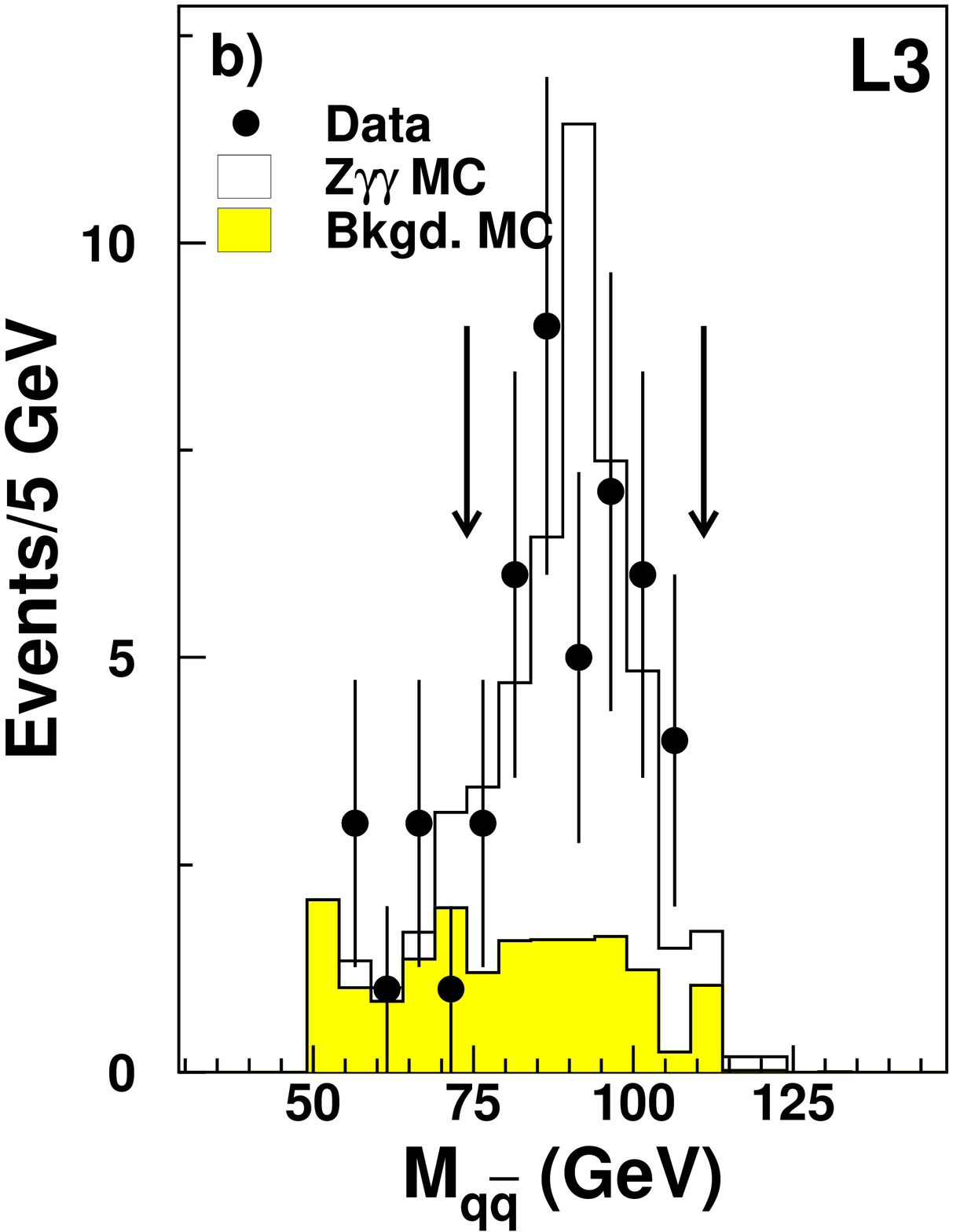}} \\
      \mbox{\includegraphics[width=.5\figwidth]{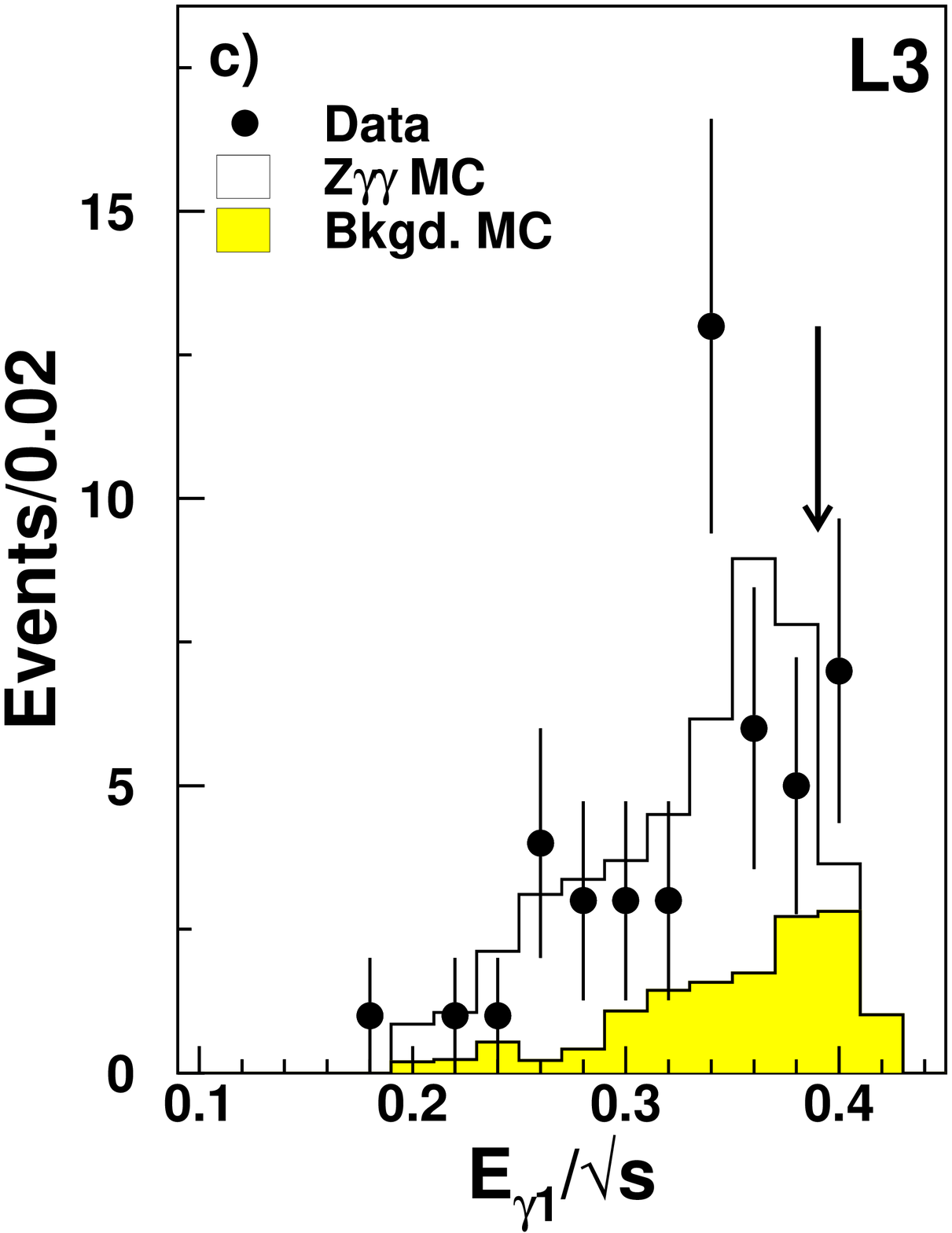}} &
      \mbox{\includegraphics[width=.5\figwidth]{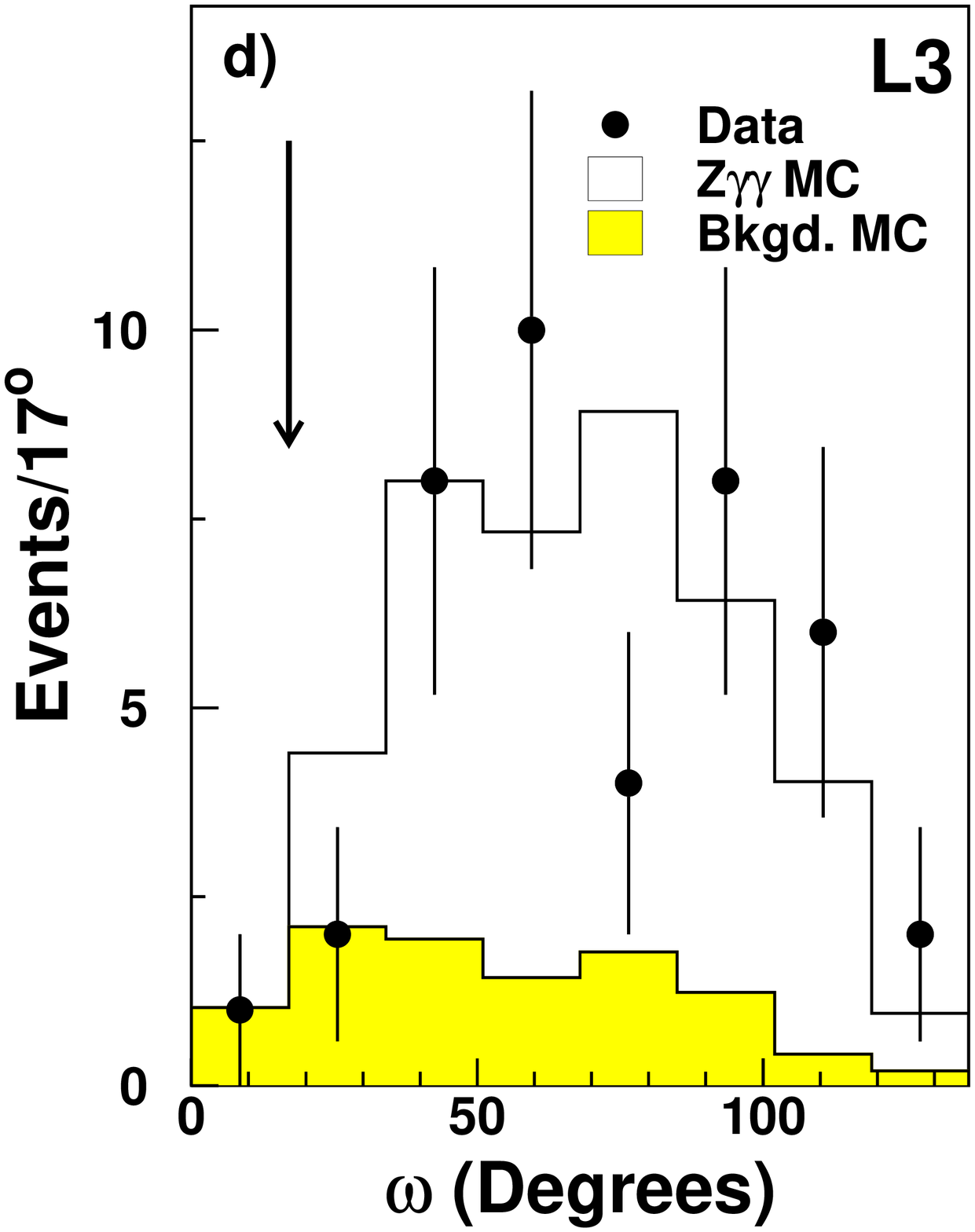}} \\
    \end{tabular}
    \icaption{Distributions of, a), the relativistic velocity
      $\beta_{\Zo}$ of the Z boson reconstructed from the measured
      photons, b), the invariant mass $M_{\qqbar}$ of the hadronic
      system, c), the scaled energy $E_{\gamma 1}/\sqrt{s}$ of the
      most energetic photon and, d), the angle $\omega$ between the
      least energetic photon and the nearest jet. Data, signal and
      background Monte Carlo samples are shown. 
      Monte Carlo predictions are normalised to the
      integrated luminosity of the data. The arrows show the positions
      of the final selection cuts.  In each plot, cuts on all other
      variables have been applied.
      \label{fig:1}}
  \end{center}
\end{figure}

Events with a single initial state radiation photon, such as those shown in
Figure 1d and Figure 1e, are rejected by an upper bound on the energy
$E_{\gamma 1}$ of the most energetic photon. This cut is chosen as
$E_{\gamma 1}<79.9\GeV$ at $\sqrt{s} = 204.8\GeV$ and $E_{\gamma
1}<80.6\GeV$ at $\sqrt{s} = 205.6\GeV$.  A lower bound of 17$^\circ$
on the angle $\omega$ between the direction of the least energetic
photon and that of the  
closest jet is also imposed.  Data and Monte Carlo distributions of
these selection variables are presented in Figure~\ref{fig:1}. Good
agreement is observed.

Table~1 lists the signal efficiencies and the numbers of events
selected in the data and Monte Carlo samples. A signal purity around 75\%
is obtained.  The dominant background  consists of hadronic
events with photons. Half of these are ``non-resonant'' events,
the other half being events with final state radiation or fake photons.

\section*{Cross Section Measurement}

\begin{figure}[p]
  \begin{center}
    \begin{tabular}{c}
      \mbox{\rotatebox{90}{
          \includegraphics[width=.73\figwidth]{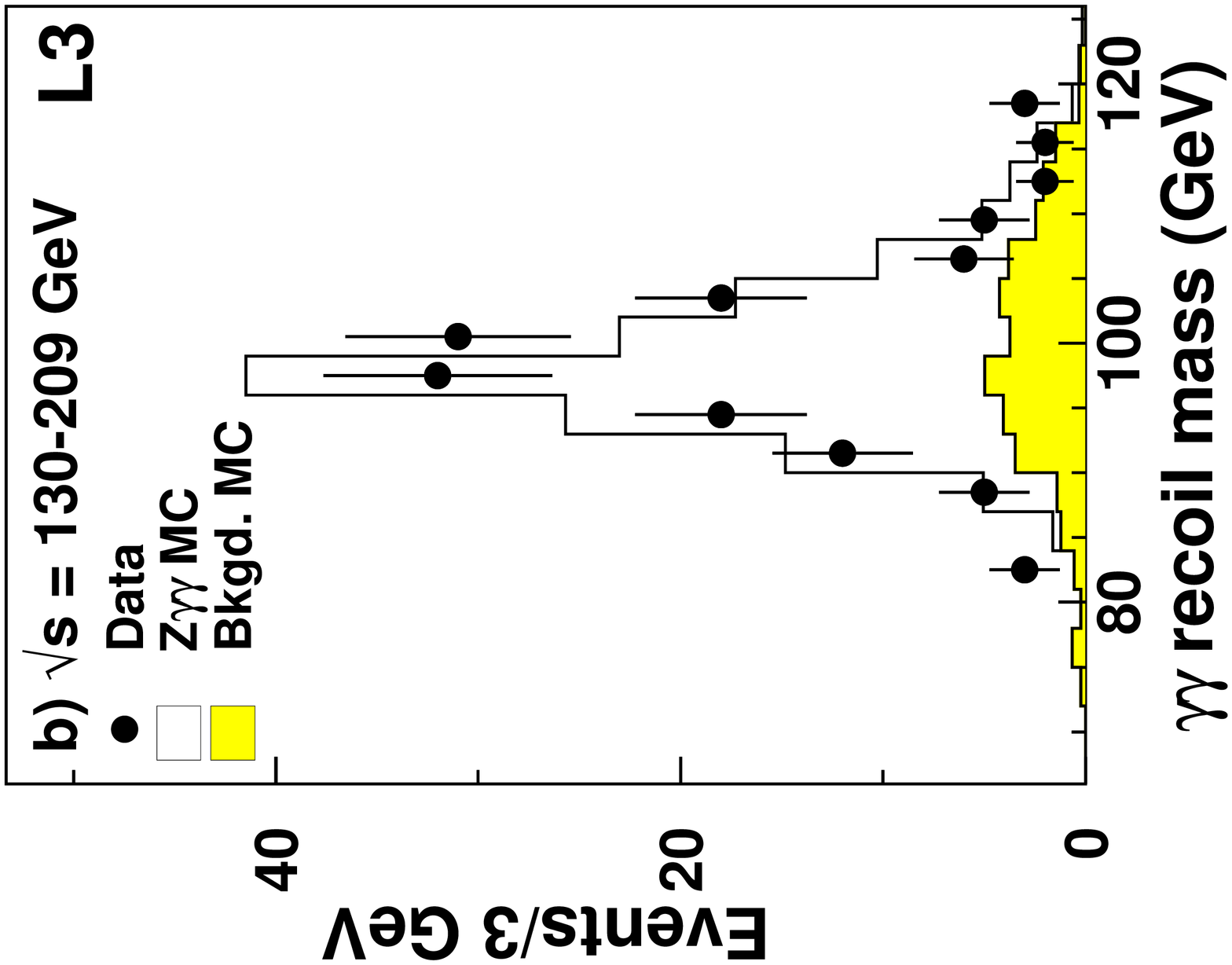}}} \\
      \mbox{\rotatebox{90}{
          \includegraphics[width=.73\figwidth]{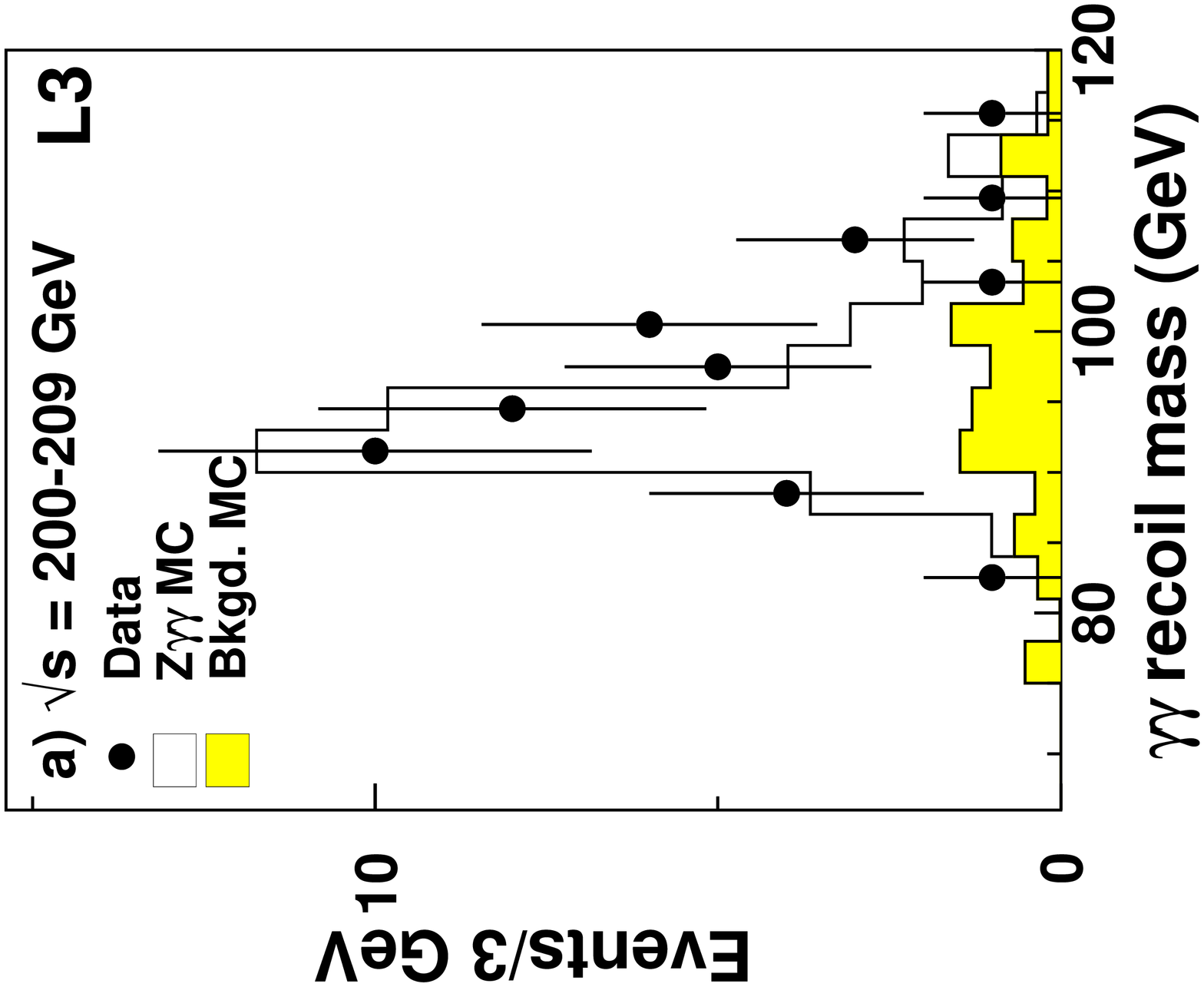}}} \\
    \end{tabular}
    \icaption{Mass recoiling from photon pairs in data, signal and
    background Monte Carlo for, a), the data sample analysed in this
    Letter and, b), the total sample collected above the Z
    resonance. Monte Carlo predictions are normalised to the
    integrated luminosity of the data.
    \label{fig:2}}
  \end{center}
\end{figure}

A clear Z signal is observed in the spectrum of the recoil
mass to the two photons, as presented in Figure~\ref{fig:2}a.  The
$\epem\ra\Zo\gamma\gamma\ra\qqbar\gamma\gamma$ cross section,
$\sigma$, is determined in the kinematical region defined by
Equation~(1) at each average $\sqrt{s}$ by a 
fit to the recoil mass spectrum. The background predictions and the
signal shape are fixed, while the signal normalisation is
fitted.  The results are\footnote{
The cross section 
is also measured in the more restrictive phase space defined by
tightening the bounds on $\theta_{\gamma}$ and $\theta_{\rm \gamma q}$
to $|\cos{\theta_{\gamma}}|< 0.95$ and $\cos{\theta_{\rm \gamma q}}<
0.9$. For the full 215\,pb$^{-1}$ at the combined average $\sqrt{s}$ of
$205.9\GeV$, the result is: $\sigma(205.9\GeV) = 0.18 \pm0.06 \pm
0.02\,{\mathrm{pb}}$, with a Standard Model expectation of
$\sigma_{SM}=0.172\pm0.003\,{\mathrm{pb}}$.
}:

\begin{center}
\begin{tabular}{cccc}
$\sigma(204.8\GeV)$ & = & $0.30^{+0
.11}_{-0.09}\pm
0.03\,{\mathrm{pb}}$ & $(\rm \sigma_{SM}=0.287\pm 0.003\,{\mathrm{pb}})$\phantom{,}\\
$\sigma(206.6\GeV)$ & =
& $0.25^{+0.07}_{-0.06}\pm
0.03\,{\mathrm{pb}}$ & $(\rm \sigma_{SM}=0.281\pm 0.003\,{\mathrm{pb}})$.\\
\end{tabular}
\end{center}
Here and below, the first quoted uncertainties are statistical and the second
ones systematic.
The systematic uncertainties on the cross section
measurement are of the order of 10\%~\cite{l3pap2}, mainly due to
the limited Monte Carlo statistics and the uncertainty on the energy scale of the detector.

The measurements are in good agreement with the theoretical
predictions, $\sigma_{\rm SM}$, as calculated with the KK2f Monte Carlo
program.  The uncertainty on the predictions (1.5$\%$) is the quadratic sum
of the theory uncertainty~\cite{KK2f} and the statistical uncertainty
of the Monte Carlo sample used for the calculation.  These results and
those obtained at lower centre-of-mass energies~\cite{l3pap2} are
compared in Figure~\ref{fig:3} to the expected Standard Model cross
section as a function of $\sqrt{s}$.

Figure~\ref{fig:2}b shows the recoil mass spectrum for the total data
sample of $\rm 712.9\,pb^{-1}$ collected at LEP above the Z resonance,
comprising the data discussed in this Letter and those at lower
centre-of-mass energies~\cite{l3pap2}. A fit to this spectrum
determines the ratio $R_{\rm \Zgg}$ between all the observed data and
the signal expectations as:
\begin{displaymath}
R_{\rm \Zgg} = \frac{\sigma}{\sigma_{\rm SM}} = 0.86 \pm 0.09 \pm 0.06,
\end{displaymath}  
in agreement with the Standard Model.  The correlation of systematic
uncertainties between the different data samples amounts to 50\% and is taken into
account in the fit.

\begin{table}[htb]
  \begin{center}
    \begin{tabular}{|c||c|c|c||c|c|c|}
\hline
\rule{0pt}{12pt} $\sqrt{s}({\rm GeV})$  & $\varepsilon(\%)$ & $\rm Data$ & ${\rm Monte~Carlo}$ 
 & $N_s$  & $N^{\rm q\bar{q}}_b$  & $N^{Other}_b$   \\
\hline
204.8 & 51 &17 &  14.7 $\pm$ 0.5 & 11.3 $\pm$ 0.5  & 3.09 $\pm$ 0.02 & 0.31 $\pm$ 0.03\\ 
206.6 & 50 &23 &  24.7 $\pm$ 0.5 & 19.5 $\pm$ 0.5  & 4.53 $\pm$ 0.04 & 0.67 $\pm$ 0.03\\
\hline
\end{tabular}
\icaption[tab:1]{Results of the
      $\epem\ra\Zo\gamma\gamma\ra\qqbar\gamma\gamma$ selection. The
      signal efficiencies, $\varepsilon$, are given, together with the
      observed and expected numbers of events. Expectations for
      signal, $N_s$, hadronic processes with photons, $N^{\rm
      q\bar{q}}_b$, and other backgrounds, $N^{Other}_b$, are
      listed. Uncertainties are due to Monte Carlo statistics.}
  \end{center}
\end{table}

\section*{Constraints on Quartic Gauge Boson Couplings}

\begin{figure}[p]
  \begin{center}
    \mbox{\includegraphics[width=\figwidth]{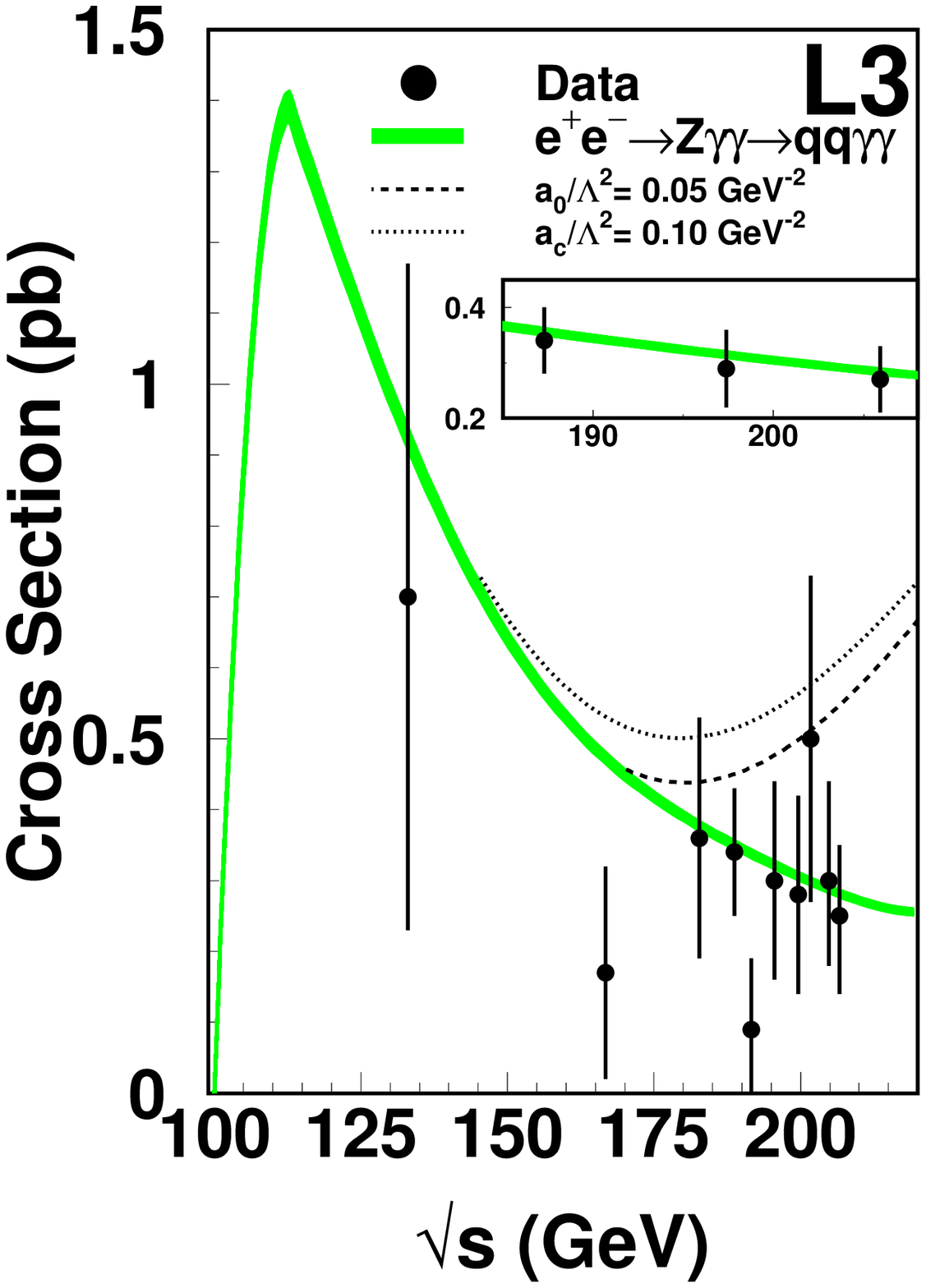}}
    \icaption{The cross section of the process
      $\epem\ra\Zo\gamma\gamma\ra\qqbar\gamma\gamma$ as a function of
      $\sqrt{s}$. The signal is defined by the phase-space cuts of
      Equation (1).  The width of the band corresponds to the
      statistical and theoretical uncertainties of the predictions of
      the KK2f Monte Carlo.  Dashed and dotted lines represent
      anomalous QGC predictions for $a_0/\Lambda^2$ = 0.05 ${\rm
      GeV^{-2}}$ and $a_c/\Lambda^2$ = 0.10 ${\rm GeV^{-2}}$,
      respectively. The inset presents three combined samples: $\rm
      231.6 pb^{-1}$ at $\sqrt{s}=182.7-188.7\GeV$, $\rm 232.9
      pb^{-1}$ at $\sqrt{s}=191.6-201.7\GeV$ and the data described in
      this Letter.
    \label{fig:3}}
  \end{center}
\end{figure}

\begin{figure}[p]
  \begin{center}
          \mbox{\includegraphics[width=1.\figwidth]{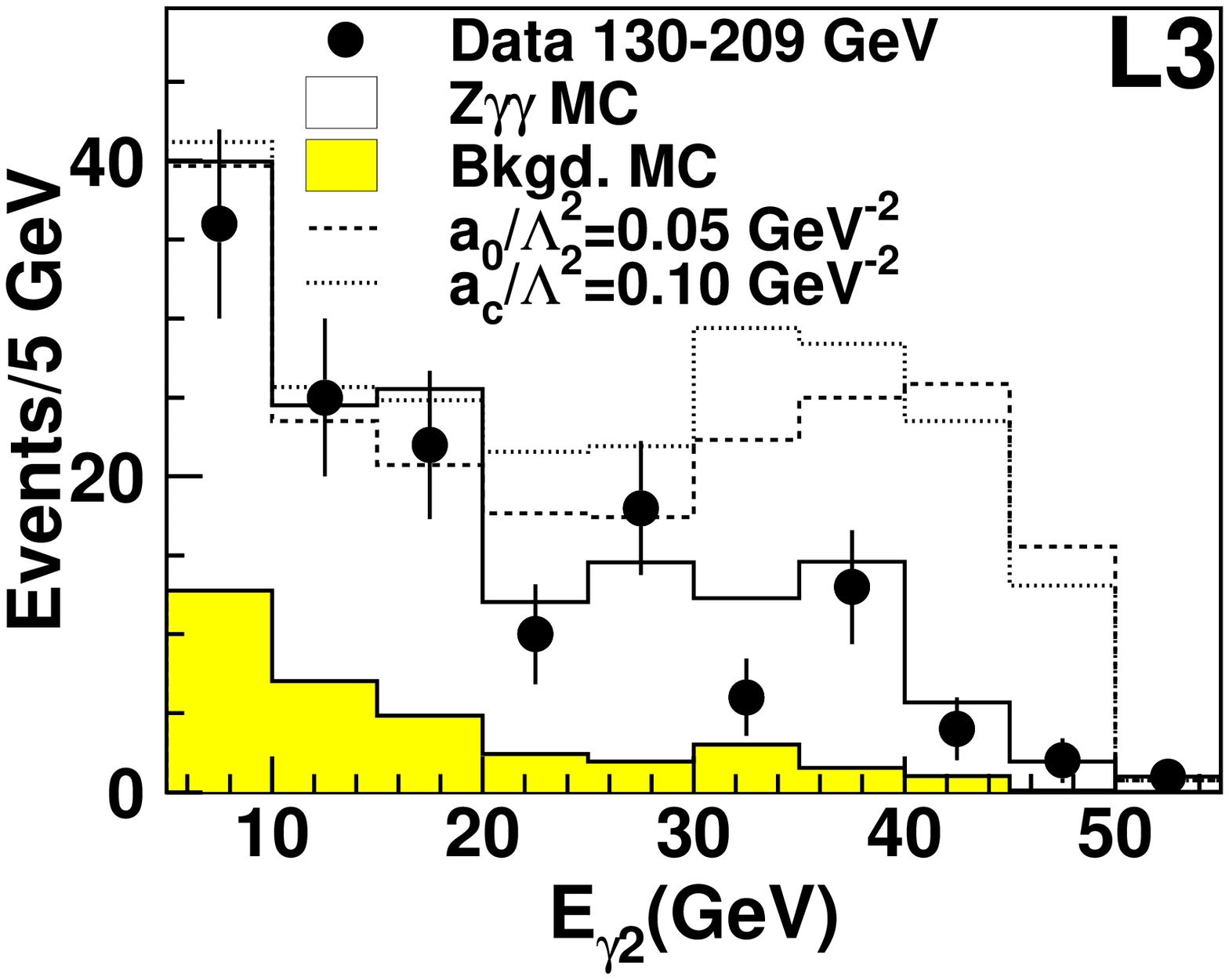}} \\
          \icaption{Energy spectrum of the least energetic photon in
          data, signal and background Monte Carlo. The full integrated
          luminosity at $\sqrt{s}=130-209\GeV$ is considered.  Monte
          Carlo predictions are normalised to the integrated
          luminosity of the data.  Examples of anomalous QGC
          predictions are also given. 
      \label{fig:dev2}}
  \end{center}
\end{figure}

Anomalous values of QGCs would manifest themselves as deviations in
the total $\epem\ra\Zo\gamma\gamma$ cross section as a function of
$\sqrt{s}$, as presented in 
Figure~\ref{fig:3}.  A harder energy spectrum for the least energetic
photon~\cite{sw} constitutes a further powerful experimental signature,
as shown in Figure \ref{fig:dev2} for the full data sample collected
at $\sqrt{s}=130-209\GeV$. QGC predictions for the cross
section and this spectrum are obtained by reweighting the Standard
Model signal Monte Carlo events. A modified
version of the WRAP~\cite{wrap} Monte Carlo program,
that includes the QGC matrix element, is used. 

\begin{figure}[hbt]
  \begin{center}
      \mbox{\includegraphics[width=1.\figwidth]{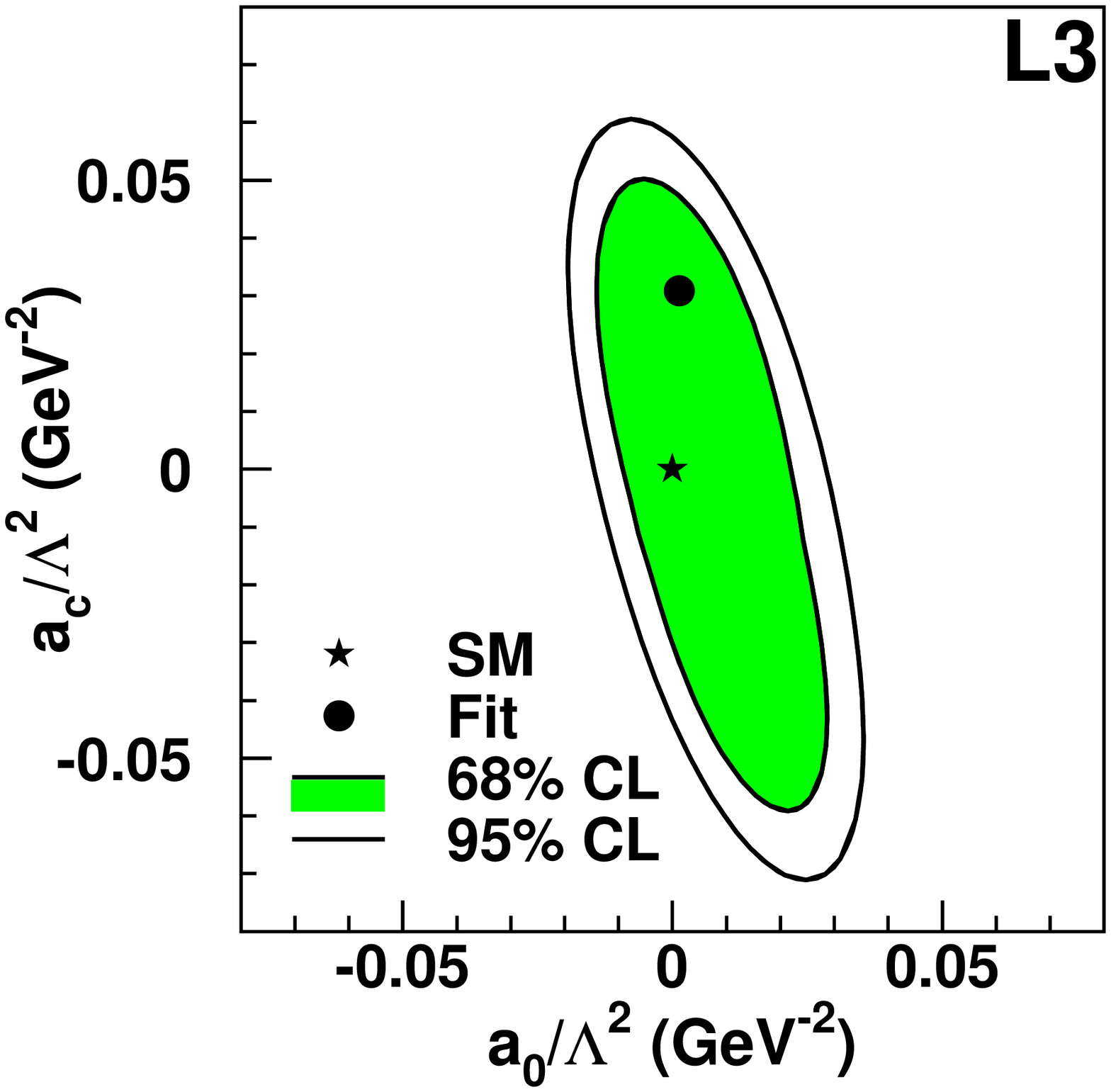}}
    \icaption{ Two dimensional confidence level contours for the
    fitted QGC parameters $a_0/\Lambda^2$ and $a_c/\Lambda^2$.  The
    fit result is shown together with the Standard Model (SM)
    predictions.
      \label{fig:5}}
  \end{center}
\end{figure}

The energy spectra of the least energetic photon are fitted for the
two $\sqrt{s}$ values discussed in this Letter and the eight values of $\sqrt{s}$ of
Reference~\citen{l3pap2}. Each of the two parameters describing the
QGCs is left free in turn, the other being fixed to zero.  The fits yield the 68\%
confidence level results:
\begin{displaymath}
a_0/\Lambda^2  =   0.00^{+ 0.02}_{-0.01} \GeV^{-2}\,\,\,{\rm and}\,\,\,
a_c/\Lambda^2  =   0.03^{+ 0.01}_{-0.02} \GeV^{-2}\,,
\end{displaymath}
in agreement with the expected Standard Model values of zero. 
A   simultaneous fit to both parameters  yields
the 95\% confidence level limits:
\begin{displaymath}
-0.02\GeV^{-2} < a_0/\Lambda^2 < 0.03 \GeV^{-2}\,\,\,{\rm and}\,\,\,
-0.07\GeV^{-2} < a_c/\Lambda^2 < 0.05 \GeV^{-2}\,,
\end{displaymath}
as shown in Figure~\ref{fig:5}.  A correlation coefficient of $-16\%$
is observed. Experimental systematic uncertainties as well as those on
the Standard Model $\epem\ra\Zo\gamma\gamma\ra\qqbar\gamma\gamma$
cross section predictions are taken into account in the fit. These
results supersede those previously obtained at lower $\sqrt{s}$~\cite{l3pap2}, as they are based on the full data sample and an improved modelling of 
QGC effects.

In conclusion, the $\epem\ra\Zo\gamma\gamma\ra\qqbar\gamma\gamma$
process is found to be well described by the Standard Model
predictions~\cite{KK2f}, with no evidence for anomalous values of QGCs.

%
%

\bibliographystyle{l3stylem}

%
%

\newpage
\typeout{   }     
\typeout{Using author list for paper 255 }
\typeout{$Modified: Jul 15 2001 by smele $}
\typeout{!!!!  This should only be used with document option a4p!!!!}
\typeout{   }
%
%
%
%
%
%

\newcount\tutecount  \tutecount=0
\def\tutenum#1{\global\advance\tutecount by 1 \xdef#1{\the\tutecount}}
\def\tute#1{$^{#1}$}
\tutenum\aachen            
\tutenum\nikhef            
\tutenum\mich              
\tutenum\lapp              
\tutenum\basel             
\tutenum\lsu               
\tutenum\beijing           
\tutenum\berlin            
\tutenum\bologna           
\tutenum\tata              
\tutenum\ne                
\tutenum\bucharest         
\tutenum\budapest          
\tutenum\mit               
\tutenum\panjab            
\tutenum\debrecen          
\tutenum\dublin            
\tutenum\florence          
\tutenum\cern              
\tutenum\wl                
\tutenum\geneva            
\tutenum\hefei             
\tutenum\lausanne          
\tutenum\lyon              
\tutenum\madrid            
\tutenum\florida           
\tutenum\milan             
\tutenum\moscow            
\tutenum\naples            
\tutenum\cyprus            
\tutenum\nymegen           
\tutenum\caltech           
\tutenum\perugia           
\tutenum\peters            
\tutenum\cmu               
\tutenum\potenza           
\tutenum\prince            
\tutenum\riverside         
\tutenum\rome              
\tutenum\salerno           
\tutenum\ucsd              
\tutenum\sofia             
\tutenum\korea             
\tutenum\purdue            
\tutenum\psinst            
\tutenum\zeuthen           
\tutenum\eth               
\tutenum\hamburg           
\tutenum\taiwan            
\tutenum\tsinghua          

{
\parskip=0pt
\noindent
{\bf The L3 Collaboration:}
\ifx\selectfont\undefined
 \baselineskip=10.8pt
 \baselineskip\baselinestretch\baselineskip
 \normalbaselineskip\baselineskip
 \ixpt
\else
 \fontsize{9}{10.8pt}\selectfont
\fi
\medskip
\tolerance=10000
\hbadness=5000
\raggedright
\hsize=162truemm\hoffset=0mm
\def\r{\rlap,}
\noindent

P.Achard\r\tute\geneva\ 
O.Adriani\r\tute{\florence}\ 
M.Aguilar-Benitez\r\tute\madrid\ 
J.Alcaraz\r\tute{\madrid,\cern}\ 
G.Alemanni\r\tute\lausanne\
J.Allaby\r\tute\cern\
A.Aloisio\r\tute\naples\ 
M.G.Alviggi\r\tute\naples\
H.Anderhub\r\tute\eth\ 
V.P.Andreev\r\tute{\lsu,\peters}\
% ...
\newpage
\begin{list}{A}{\itemsep=0pt plus 0pt minus 0pt\parsep=0pt plus 0pt minus 0pt
                \topsep=0pt plus 0pt minus 0pt}
\item[\aachen]
 I. Physikalisches Institut, RWTH, D-52056 Aachen, FRG$^{\S}$\\
 III. Physikalisches Institut, RWTH, D-52056 Aachen, FRG$^{\S}$
\item[\nikhef] National Institute for High Energy Physics, NIKHEF, 
     and University of Amsterdam, NL-1009 DB Amsterdam, The Netherlands
\item[\mich] University of Michigan, Ann Arbor, MI 48109, USA
\item[\lapp] Laboratoire d'Annecy-le-Vieux de Physique des Particules, 
     LAPP,IN2P3-CNRS, BP 110, F-74941 Annecy-le-Vieux CEDEX, France
\item[\basel] Institute of Physics, University of Basel, CH-4056 Basel,
     Switzerland
\item[\lsu] Louisiana State University, Baton Rouge, LA 70803, USA
\item[\beijing] Institute of High Energy Physics, IHEP, 
  100039 Beijing, China$^{\triangle}$ 
\item[\berlin] Humboldt University, D-10099 Berlin, FRG$^{\S}$
\item[\bologna] University of Bologna and INFN-Sezione di Bologna, 
     I-40126 Bologna, Italy
\item[\tata] Tata Institute of Fundamental Research, Mumbai (Bombay) 400 005, India
\item[\ne] Northeastern University, Boston, MA 02115, USA
\item[\bucharest] Institute of Atomic Physics and University of Bucharest,
     R-76900 Bucharest, Romania
\item[\budapest] Central Research Institute for Physics of the 
     Hungarian Academy of Sciences, H-1525 Budapest 114, Hungary$^{\ddag}$
\item[\mit] Massachusetts Institute of Technology, Cambridge, MA 02139, USA
\item[\panjab] Panjab University, Chandigarh 160 014, India.
\item[\debrecen] KLTE-ATOMKI, H-4010 Debrecen, Hungary$^\P$
\item[\dublin] Department of Experimental Physics,
  University College Dublin, Belfield, Dublin 4, Ireland
\item[\florence] INFN Sezione di Firenze and University of Florence, 
     I-50125 Florence, Italy
\item[\cern] European Laboratory for Particle Physics, CERN, 
     CH-1211 Geneva 23, Switzerland
\item[\wl] World Laboratory, FBLJA  Project, CH-1211 Geneva 23, Switzerland
\item[\geneva] University of Geneva, CH-1211 Geneva 4, Switzerland
\item[\hefei] Chinese University of Science and Technology, USTC,
      Hefei, Anhui 230 029, China$^{\triangle}$
\item[\lausanne] University of Lausanne, CH-1015 Lausanne, Switzerland
\item[\lyon] Institut de Physique Nucl\'eaire de Lyon, 
     IN2P3-CNRS,Universit\'e Claude Bernard, 
     F-69622 Villeurbanne, France
\item[\madrid] Centro de Investigaciones Energ{\'e}ticas, 
     Medioambientales y Tecnol\'ogicas, CIEMAT, E-28040 Madrid,
     Spain${\flat}$ 
\item[\florida] Florida Institute of Technology, Melbourne, FL 32901, USA
\item[\milan] INFN-Sezione di Milano, I-20133 Milan, Italy
\item[\moscow] Institute of Theoretical and Experimental Physics, ITEP, 
     Moscow, Russia
\item[\naples] INFN-Sezione di Napoli and University of Naples, 
     I-80125 Naples, Italy
\item[\cyprus] Department of Physics, University of Cyprus,
     Nicosia, Cyprus
\item[\nymegen] University of Nijmegen and NIKHEF, 
     NL-6525 ED Nijmegen, The Netherlands
\item[\caltech] California Institute of Technology, Pasadena, CA 91125, USA
\item[\perugia] INFN-Sezione di Perugia and Universit\`a Degli 
     Studi di Perugia, I-06100 Perugia, Italy   
\item[\peters] Nuclear Physics Institute, St. Petersburg, Russia
\item[\cmu] Carnegie Mellon University, Pittsburgh, PA 15213, USA
\item[\potenza] INFN-Sezione di Napoli and University of Potenza, 
     I-85100 Potenza, Italy
\item[\prince] Princeton University, Princeton, NJ 08544, USA
\item[\riverside] University of Californa, Riverside, CA 92521, USA
\item[\rome] INFN-Sezione di Roma and University of Rome, ``La Sapienza",
     I-00185 Rome, Italy
\item[\salerno] University and INFN, Salerno, I-84100 Salerno, Italy
\item[\ucsd] University of California, San Diego, CA 92093, USA
\item[\sofia] Bulgarian Academy of Sciences, Central Lab.~of 
     Mechatronics and Instrumentation, BU-1113 Sofia, Bulgaria
\item[\korea]  The Center for High Energy Physics, 
     Kyungpook National University, 702-701 Taegu, Republic of Korea
\item[\purdue] Purdue University, West Lafayette, IN 47907, USA
\item[\psinst] Paul Scherrer Institut, PSI, CH-5232 Villigen, Switzerland
\item[\zeuthen] DESY, D-15738 Zeuthen, 
     FRG
\item[\eth] Eidgen\"ossische Technische Hochschule, ETH Z\"urich,
     CH-8093 Z\"urich, Switzerland
\item[\hamburg] University of Hamburg, D-22761 Hamburg, FRG
\item[\taiwan] National Central University, Chung-Li, Taiwan, China
\item[\tsinghua] Department of Physics, National Tsing Hua University,
      Taiwan, China
\item[\S]  Supported by the German Bundesministerium 
        f\"ur Bildung, Wissenschaft, Forschung und Technologie
\item[\ddag] Supported by the Hungarian OTKA fund under contract
numbers T019181, F023259 and T037350.
\item[\P] Also supported by the Hungarian OTKA fund under contract
  number T026178.
\item[$\flat$] Supported also by the Comisi\'on Interministerial de Ciencia y 
        Tecnolog{\'\i}a.
\item[$\sharp$] Also supported by CONICET and Universidad Nacional de La Plata,
        CC 67, 1900 La Plata, Argentina.
\item[$\triangle$] Supported by the National Natural Science
  Foundation of China.
\end{list}
}
\vfill


\newpage

%
%

\end{document}